\documentclass[10pt,journal]{IEEEtran}

\ifCLASSINFOpdf
\else
   \usepackage[dvips]{graphicx}
\fi
\usepackage{url}
\usepackage{graphicx}

\usepackage{subfigure}

\usepackage{verbatim}
\usepackage{multirow}
\usepackage{array}
 \usepackage{cite}
\usepackage{amsmath}
\usepackage{algorithm}
\usepackage{algorithmic}
\usepackage{multirow}
\usepackage{comment}
\usepackage[justification=centering]{caption}

\usepackage{xcolor}

\usepackage{array}
\newcommand{\PreserveBackslash}[1]{\let\temp=\\#1\let\\=\temp}
\newcolumntype{C}[1]{>{\PreserveBackslash\centering}p{#1}}
\newcolumntype{R}[1]{>{\PreserveBackslash\raggedleft}p{#1}}
\newcolumntype{L}[1]{>{\PreserveBackslash\raggedright}p{#1}}
\begin{document}

\title{Adversarial Feature Learning and Unsupervised Clustering based Speech Synthesis for Found Data with Acoustic and Textual Noise}

\author{Shan Yang\thanks{Shan Yang and Lei Xie are with the Audio, Speech and Language Processing Group, Northwestern Polytechnical University, China.}, \IEEEmembership{Student Member, IEEE}, Yuxuan Wang\thanks{Yuxuan Wang is with Bytedance AI Lab, Mountain View, USA}, \\ Lei Xie, \IEEEmembership{Senior Member, IEEE}}

\maketitle

\begin{abstract}
Attention-based sequence-to-sequence (seq2seq) speech synthesis has achieved extraordinary performance. But a studio-quality corpus with manual transcription is necessary to train such seq2seq systems. In this paper, we propose an approach to build high-quality and stable seq2seq based speech synthesis system using challenging found data, where training speech contains noisy interferences (acoustic noise) and texts are imperfect speech recognition transcripts (textual noise). To deal with text-side noise, we propose a VQVAE based heuristic method to compensate erroneous linguistic feature with phonetic information learned directly from speech. As for the speech-side noise, we propose to learn a noise-independent feature in the auto-regressive decoder through adversarial training and data augmentation, which does not need an extra speech enhancement model. Experiments show the effectiveness of the proposed approach in dealing with text-side and speech-side noise. Surpassing the denoising approach based on a state-of-the-art speech enhancement model, our system built on noisy found data can synthesize clean and high-quality speech with MOS close to the system built on the clean counterpart.
\end{abstract}


\IEEEpeerreviewmaketitle

\vspace{-10pt}
\section{Introduction}
\label{sec:intro}
\IEEEPARstart{R}{ecently}, text-to-speech (TTS) has been significantly advanced with the wide use of deep neural networks (DNN). With the success of attention-based sequence-to-sequence (seq2seq) approach in machine translation~\cite{bahdanau2014neural,sutskever2014sequence}, DNN based speech synthesis has evolved into an \textit{end-to-end} (E2E) framework, which unifies acoustic and duration modeling in a compact seq2seq paradigm, discarding frame-wise linguistic-acoustic mapping~\cite{wang2017tacotron,shen2017natural,tachibana2017efficiently,li2019close,yang2020localness}. 
To achieve the best performance from a seq2seq system, studio-quality speech recordings with manual transcripts are necessary.  Leveraging huge amount of speech resources available in public domain, or so-called \textit{found data}, has drawn much interests lately. However, it is challenging to build a TTS system on low-quality found data as 1) speech may be contaminated by channel and environmental noises -- \textit{acoustic noise} and 2) transcripts generated by an automatic speech recognizer (ASR) contain inevitable errors -- \textit{textual noise}.

There are several recent studies addressing acoustic noise for seq2seq-based E2E TTS. A straightforward idea is to use de-noised audio from speech enhancement to build the acoustic model~\cite{valentini2016investigating}. But the inevitable distortion on training speech will propagate to the synthesized speech, resulting in clear quality deterioration. This conclusion has been further confirmed by another unsupervised source separation approach~\cite{gurunath2019disentangling}, where multi-node variational auto-encoder (VAE) was introduced to remove background music from the found speech for speech synthesis. The unstable separation directly affects the TTS quality. Another solution is to disentangle the speaker and noise attributes directly in the speech synthesis model. The approach in~\cite{hsu2019disentangling} first encodes the reference audio to disentangle speaker and noise, where adversarial factorization is used to encourage such disentanglement, and then inject the encodings into an acoustic decoder to produce clean speech. This approach can generate clean speech with the help of a clean reference audio, but there is a strong assumption to conduct domain adversarial training-- the audio of one speaker has a fixed type of acoustic noise, which greatly limits its application. In more practical situation, recording conditions may vary and the collected data for the target speaker may come from different sources with different noise interferences. 
 
We only find one recent study investigating the textual noise. In~\cite{fong2019investigating}, the robustness of E2E systems to textual noise has been studied by manually corrupting text and using erronous ASR transcripts. Results suggest that E2E systems only partially robust to training on imperfectly-transcribed data, and substitutions and deletions pose a serious problem. To the best our knowledge, there is still no solution to deal with textual noise in E2E TTS. Moreover, in many circumstances, building TTS systems on noisy found data has to deal with both textual and acoustic noise simultaneously -- noisy speech transcribed by an ASR system with text errors.

This paper addresses both acoustic and textual noise interferences for building seq2seq-based speech synthesis system on noisy found data. To deal with textual noise, we propose a heuristic method to compensate the erroneous linguistic feature with phonetic information learned directly from speech. Specifically, VQVAE-based unsupervised clustering on the training speech is adopted to obtain latent phonetic representation, which is combined with the context vector from the text encoder output to produce synthesized speech. As for the acoustic noise, we propose to learn a noise-independent feature in the auto-regressive decoder through adversarial training and data augmentation, which does not need an extra speech enhancement model or strong assumption for noise conditions in~\cite{hsu2019disentangling}. Specifically, with the help of the clean data from another speaker, we adopt a domain classification network with a gradient reversal layer in the auto-regressive decoder to disentangle the noise conditions in latent feature space. Experiments show the effectiveness of the proposed approaches in dealing with textual and acoustic noise. Surpassing the denoising approach based on a state-of-the-art speech enhancement model, our system built on noisy data (speech SNR = 4dB, text CER=23.3\%) can synthesize clean and high-quality speech with MOS close to the system built on the clean counterpart.

\section{Proposed methods for found data}
\label{sec:method}
\begin{figure}[th]
        \centering
        \includegraphics[width=1.0\linewidth]{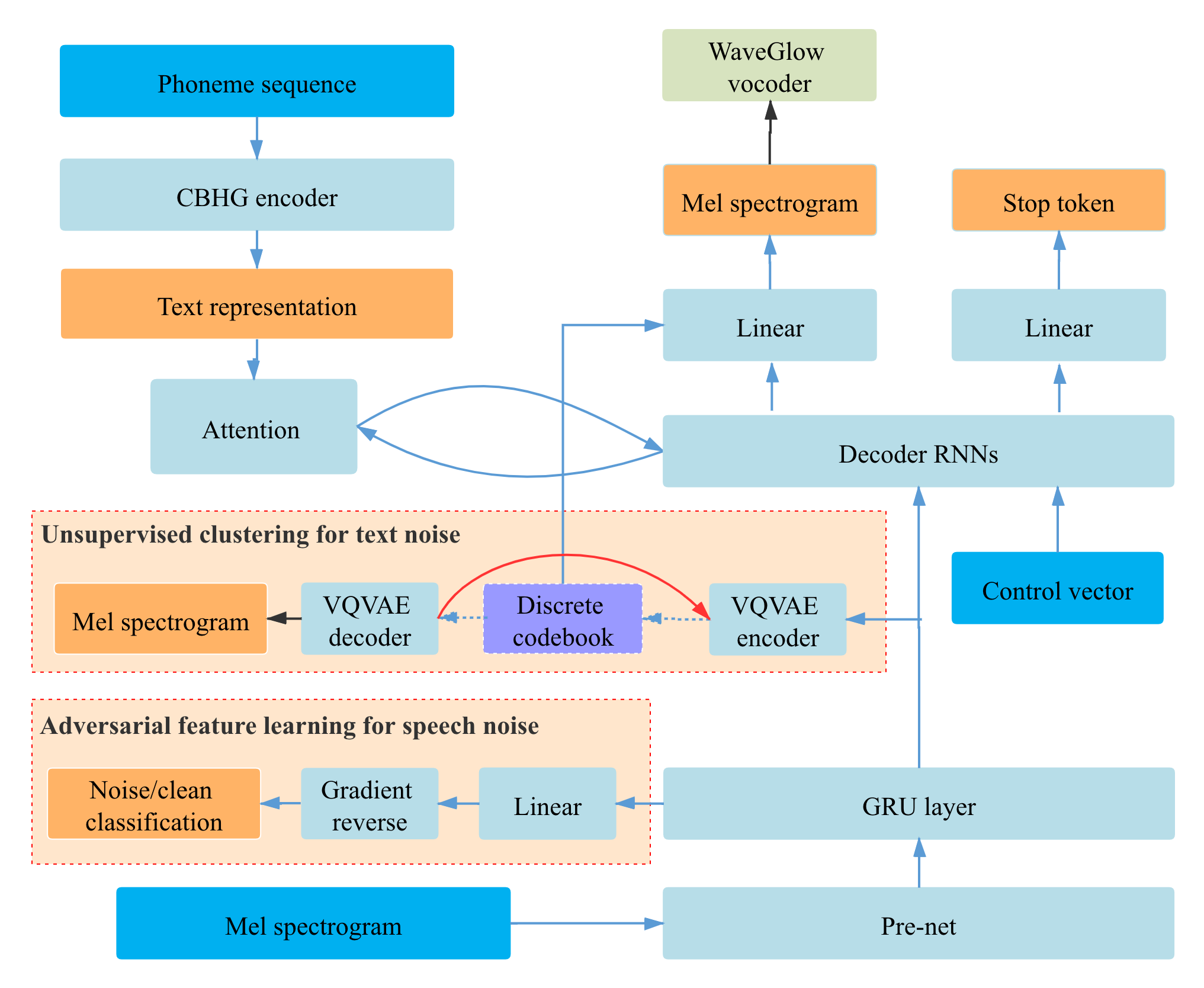}
        \caption{Proposed methods for found data}
        \vspace{-10pt}
        \label{fig:framework}
\end{figure}

Fig.~\ref{fig:framework} illustrates our seq2seq-based speech synthesis framework, which shares the similar architecture with Tacotron~\cite{wang2017tacotron,shen2017natural}. It is composed of a CBHG-based text encoder~\cite{wang2017tacotron} and an auto-regressive decoder for mel-spectrogram generation, while attention mechanism serves as the bridge. The WaveGlow~\cite{prenger2019waveglow} vocoder is used to reconstruct the waveforms from Mel-spectrogram. Our approaches dealing with text and speech noise are built on this baseline system. Below we briefly describe the seq2seq-based speech synthesis.

For seq2seq TTS framework, suppose each speech utterance with $M$ frames of acoustic features $\mathbf{y}=(y_1, y_2, ..., y_M)$ has corresponding $N$ frames of golden character- or phoneme-level transcript $\mathbf{x}=(x_1, x_2, ..., x_N)$. The goal is to maximize the log probability $P(\mathbf{y}|\mathbf{x})$. And in the basic attention-based seq2seq framework~\cite{wang2017tacotron}, the decoder output $\hat y_t$ is computed from
\begin{equation}
  \hat y_t = f(y_{t-1}, c_t) \quad \mbox{where $c_t=g(y_{t-1}, \mathbf{x})$},
  \label{eq2}
\end{equation}
where $y_{t-1}$ is the ground-truth acoustic frame at time $t-1$, and $c_t$ is the context vector computed from the attention function $g(\cdot)$, which includes a content- or non-content based score function to measure the contribution of each memory $x_i$~\cite{bahdanau2014neural,chorowski2015attention,battenberg2019location}. The objective function is to minimize the distance between predicted $\mathbf{\hat y}$ and target $\mathbf{y}$.

\vspace{-7pt}
\subsection{Unsupervised clustering for textual noise}
In the found data scenario, the golden transcript $\mathbf{x}$ is unavailable for model training. Thus we need an extra speech recognizer conducting auto-transcription to get $\mathbf{\hat x}=(\hat x_1, \hat x_2, ..., \hat x_K)$. Compared to the reference $\mathbf{x}$, $\mathbf{\hat x}$ may have irregular insertion, deletion or substitution errors. So the key problem is how to model the relations between unmatched speech $\mathbf{y}$ and text $\mathbf{\hat x}$. As described in Eq~\eqref{eq2}, given a previous speech frame $y_{t-1}$, the attention mechanism computes the contribution of each $\hat x_i$ in hidden space to generate $\hat y_t$. Due to the recognition error of ASR and the monotonic nature of speech generation, the speech $y_t$ may focus on the unrelated $\hat x_i$ rather than the correct $x_i$, which mostly causes the mispronunciation problem according to our experiments.  

It's almost impossible to directly handle the text noise in the speech synthesis task as the supervision labels for text and speech are totally unavailable. Our approach dealing with text noise is motivated by recent works on unsupervised speech unit discovery, which has shown that phoneme-like clusterings can be automatically learned from speech in an unsupervised manner~\cite{van2017neural,dunbar2019zero}. Specifically, similar latent features from waveforms tend to be categorized into different clusterings that act as a high-level speech descriptors closely related to phonemes~\cite{van2017neural}. To deal with text noise in found-data TTS, we propose a heuristic approach to conduct unsupervised clustering in the auto-regressive decoder to guide the speech generation with learnable latent phoneme representation, as shown in the unsupervised clustering module in Fig.~\ref{fig:framework}.


In details, the context vector in the basic seq2seq framework is only computed from the output of text encoder, which inevitably contains text noise due to the inaccurate speech recognizer. In the proposed method, we compensate such errors with phonetic representation learned directly from speech. The context vector and the phonetic latent features are both injected to the decoder to produce synthesized speech, reducing the mismatch between speech and noisy text. There are several ways to learn the above discrete phonetic space~\cite{van2017neural, dilokthanakul2016deep}. In our work, we adopt vector quantized variational autoencoder (VQVAE) to obtain a learnable discrete clusterings space $e$, which we assume is related to phoneme-like units~\cite{van2017neural}. 

Along with the basic auto-regressive process, the latent representation of $y$ is also fed into the VQVAE encoder to obtain latent $z_e(y)$. we can obtain the discrete latent feature $z_q(y) \in e$ through
\begin{equation}
  z_q(y) = e_k,\quad \mbox{where $k = \arg \min_i{||z_e(y)-e_i||_2} $}.
  \label{vq2}
\end{equation} 
Here $z_q(y)$ is treated as the latent phoneme representation clustered from speech and can be utilized to reconstruct back to speech through the VQVAE decoder. 

Besides, the selected latent clustering $z_q(y)$ is also fed into the auto-regressive decoder along with the context vector $c_t$. Therefore Eq.~\eqref{eq2} is updated to
\begin{equation}
  \hat y_t = f(y_{t-1}, c_t, z_q(y_{t-1})).
  \label{vq3}
\end{equation}

The objective function of the whole network is:
\begin{eqnarray}
     Loss & = & Loss_{recons} + ||sg[z_e(y)] - e||_2 \nonumber \\
     & & + \alpha * ||z_e(y) - sg[e]||_2
\label{text.loss}
\end{eqnarray}
where $Loss_{recons}$ includes the reconstruction loss of both auto-regressive decoder and the VQVAE model, and $sg[\cdot]$ is a stop-gradient operator which has zero partial derivatives at the $\cdot$ operation. $\alpha$ is the weight of the commitment loss to make sure the encoder commits to an embedding~\cite{van2017neural}. Since there is no real gradient defined for Eq.~\eqref{vq2}, we copy the gradient from VQVAE decoder input to the encoder output, as shown in the red line in Fig.~\ref{fig:framework}.

\vspace{-7pt}
\subsection{Adversarial feature learning for acoustic noise}
Speech utterance $\mathbf{y}$ in found data may contain different types of background noise, which directly affects the performance of attention function $g(\cdot)$ and the whole model. We can apply an external speech enhancement module to obtain de-noised speech feature $\mathbf{\overline y}$ from $\mathbf{y}$ for downstream speech synthesis model training, but it may cause distortion problem in the generated speech~\cite{valentini2016investigating,gurunath2019disentangling} . 

In order to mitigate the negative effects from speech noise in $\mathbf{y}$, we propose to use adversarial training to obtain the noise-independent latent feature $z_s=G_{adv}(\mathbf{y})$, where $G_{adv}(\cdot)$ is the proposed adversarial module. As shown in Fig.~\ref{fig:framework}, the adversarial module contains a pre-net, a single unidirectional gated recurrent unit (GRU) network, and a classification network with a gradient reverse layer (GRL)~\cite{sun2017unsupervised}. The classification task is designed to classify the speech sample into clean/noisy. Here, since we do no have clean samples, similar to the data augmentation strategy in~\cite{hsu2019disentangling}, we use another clean speech dataset along with the noisy samples to train the classification network. For a common classification network, the logistics of the last latent layer often represent the classification information (noise/clean condition in this work). When conducting the gradient reverse operation, its aim becomes disentangling the noise information to obtain the noise-independent features~\cite{sun2017unsupervised}, or encouraging $z_s$ not to be informative about the acoustic condition (noisy or clean). In ~\cite{hsu2019disentangling}, GRL is also adopted to disentangle noise from a reference audio to control the condition of speech synthesis. But we learn the noise-independent features $z_s$ directly from the input speech $\mathbf{y}$. Therefore, in the speech synthesis stage, we do not need a clean reference audio to generate clean speech.  With the GRL, the context vector $c_t$ in Eq.~\eqref{eq2} becomes

\begin{equation}
  c_t=g(G_{adv}(y_{t-1}), \mathbf{x}).
  \label{eq4}
\end{equation}

Since there is an extra classification network, the final objective function is
\begin{eqnarray}
     Loss & = & Mel_{rmse} + \beta * Noise_{ce} ,
\label{e.loss}
\end{eqnarray}
where $Mel_{rmse}$ denotes the Mel-spectrogram reconstruction loss, $Noise_{ce}$ is the cross entropy loss for noise classification, and $\beta$ is the weight for the classification loss.

\vspace{-7pt}
\section{Experiments}
\label{sec:exp}
\vspace{-1pt}
In our experiments, we use an open-source Chinese corpus, which contains 10 hours speech of a female speaker. To obtain the target noisy dataset, we mix the clean speech with random types of noises from the CHiME-4 challenge~\cite{vincent2017analysis}. We use a speech recognition module to transcribe the noisy speech, where the character error rate (CER) depends on the signal-to-noise ratio (SNR). In order to do the adversarial training, we use another clean corpus with 11 hours speech from another Chinese female speaker as the \textit{clean} data. Another copy of this corpus is mixed with random noise from CHiME-4, together with the target noisy corpus above to form the \textit{noisy} data. We use an internal speech recognition system to obtain transcripts. The CER is 8.9\% for the clean target speaker. As for the speech enhancement baselines, we test an unsupervised model Separabl~\cite{gurunath2019disentangling} and a state-of-the-art supervised model DCUnet~\cite{choi2019phase}. The Perceptual Evaluation of Speech Quality (PESQ) for Separabl and DCUnet are 2.02 and 3.00. 

We mainly analyze the phone, tone and prosody information through our text analysis module to obtain text representation. For the speech representation, 80-band mel-scale spectrogram is treated as $y$ for attention based decoder. For evaluation, we reserve 400 sentences from the target corpus to conduct objective and subjective testing. There are 20 listeners attending the mean opinion score (MOS) test as subjective evaluation\footnote{Samples can be found at \url{https://syang1993.github.io/found-data}}. Since the length of predicted acoustic features is different from the target one, we conduct dynamic time warping (DTW) to align the two sequences and then compute the mel-cepstral distortion (MCD) for objective evaluation.

\vspace{-1pt}
\subsection{Model details}
\subsubsection{Basic architecture}
For the basic seq2seq system, we follow the architecture of Tacotron and Tacotron2~\cite{wang2017tacotron,shen2017natural}. In the encoder, we adopt three feed-forward layers as pre-net followed by a CBHG module~\cite{wang2017tacotron}. As for the decoder, the acoustic feature $y$ is firstly fed into the decoder pre-net. And a unidirectional LSTM with GMM based attention mechanism~\cite{battenberg2019location} is adopted on the latent features. The basic architecture is applied to the baseline systems for noisy found data and the topline system for the original clean recordings.

\subsubsection{Unsupervised clustering}
In the proposed unsupervised clustering for dealing with textual noise, the output of decoder pre-net is fed into the VQVAE encoder, which contains two layers feed-forward networks with 256 units followed by ReLu activation. The VQVAE decoder shares the similar architecture with the above encoder. For the vector quantization module, there are 256 code vectors with 128 dimensions in the code book. The weight of commit loss is set to 0.25.
\subsubsection{Adversarial feature learning}
For the noise-independent feature learning to deal with the acoustic noise, we add a 256-unit unidirectional GRU layer on the top of pre-net in the decoder. The output of GRU layer is fed into the following decoder LSTM layer with controllable speaker and noise condition, as well as the classification network.

\vspace{-7pt}
\subsection{Experimental results}
\subsubsection{Basic systems}
We first evaluate the effects of textual and acoustic noise in training data on the baseline systems. 
Table~\ref{tab:baseline} shows the objective and subjective results, where CGER means the character-level generation error. There are 7338 Chinese characters in total in the 400 test sentences.

\begin{table}
    \centering
    \caption{The performance of basic architecture for different types of found data, where R means recordings.}
    \label{tab:baseline}
    \begin{tabular}{c|c| c| c|| c| c| c}
        \hline
        Index  & CER (\%)         & SNR  & ATT    & MOS      & MCD &CGER (\%)\\ 
       \hline
        R    &-    & -  & - & 4.41& -  & -\\
        A    &0    & clean  & GMM & 4.21& 3.08 & 0.05\\ \hline
        B    &8.8 & clean  & GMM & 3.62& 4.29 & 0.07\\  
		C    &11.7  & clean & GMM & 3.51& 4.34 & 0.10\\ 
		D    &23.3  & clean & GMM & 3.04& 4.55 & 3.24\\ 
		E    &23.3   & clean  & LSA & 2.63& 4.63& 9.69\\\hline 
        F    &0    & 8 dB  &GMM  & 2.10& 7.16 & 0.05\\
        G    &0    & 4 dB &GMM   & 1.79& 8.78 & 0.04\\\hline
        H    &23.3  & 4 dB  &GMM  & 0.78 & - & -\\
        \hline
    \end{tabular}
    \vspace{-10pt}
\end{table}

System A is the topline trained using golden transcripts and clean speech. We use System B to E to examine the test noise only, while using clean speech data to train the model. Compared to the topline, System B to D get worse as the WER increases in both objective and subjective tests, which confirms the negative effects of textual noise. Comparing System A, B and C, although the text noise affects both objective and subjective results, we find the seq2seq based model has few pronunciation errors when CER$<10\%$. Since there is no punctuation in the noisy ASR transcription, the prosody of generated speech of system B and C is unsatisfactory, which causes the worse MOS values. This can be further solved through a more robust prosody model. For more noisy text, the generated speech of System D suffers from the mispronunciation problem. Since the context vector $c_t$ directly depends on the attention alignment, we also compare the content and non-content score function. In System E, we conduct content-based location sensitive attention (LSA)~\cite{chorowski2015attention}, where text memory is taken into account. System D with non-content-based GMM attention outperforms the LSA-based System E. This is because noisy text is used to obtain the alignments in LSA, which affects the attention accuracy.

As for the acoustic noise, we use the golden transcripts to build System F and G with noisy training speech at 8dB and 4dB SNR, respectively. It's obvious that System F outperforms System G since the speech data used in F contains less noise. But the synthesized speech of both systems is noisy. System H represents the real found data condition without correct transcription and clean speech, which achieves the lowest MOS of 0.78. Note that we do not have MCD for system H, since it always crashes during generation.

\subsubsection{Unsupervised clustering}
To overcome the textual noise, we then build a system with proposed unsupervised clustering to mitigate the mispronunciation problem caused by wrong transcriptions. Table~\ref{tab:text-side} shows the performance, where VQVAE based unsupervised clustering is used in the topline System A and baseline System D. Table~\ref{tab:text-side} shows that the proposed VQVAE\_D significantly outperforms System D in both subjective and objective metrics. The proposed method decreases the character generation error rate from 3.24\% to 1.29\%. It's because that each output $\hat y_t$ also depends on the unsupervisedly discovered units from speech, which can mitigate the textual noise during training. Besides, comparing System A with VQVAE\_A, we find that unsupervised clustering will not degrade the performance of the topline system with golden transcription. 

\begin{table}
    \centering
        \caption{The performance for individual textual  noise.}
    \label{tab:text-side}
    \begin{tabular}{c| c|| c| c| c}
        \hline
        Index  & CER (\%)       & MOS       & MCD & CGER (\%)\\ 
       \hline
        A    &0    & 4.21&  3.08 & 0.05\\
        VQVAE\_A    &0    & 4.25&  3.10 & 0.06\\\hline
        D    &23.3    & 3.04&  4.55 & 3.24\\
		VQVAE\_D    &23.3    & 3.47& 4.42 & 1.29\\ 
        \hline
    \end{tabular}
    \vspace{-10pt}
\end{table}

\subsubsection{Adversarial feature clustering}
In real applications, we may have to deal with both acoustic and textual noises. Here we use adversarial feature clustering to improve System H. The performances of different approaches are summarized in Table~\ref{tab:founddata}.  We can see that the two systems that use speech enhancement to remove noises before TTS model training can obviously improve TTS performance. In System Separabl, we do not need external data for speech enhancement model training, where the de-noised speech is directly obtained through the pre-trained unsupervised multi-node VAE model from noisy data~\cite{gurunath2019disentangling}. For System DCUnet, we need extra multi-speaker speech data and noise data to train the speech enhancement model. We notice that the proposed adversarial feature learning method significantly outperforms the speech enhancement methods. System Separabl indeed decreases the noise interference in the generated speech, but the performance of such unsupervised speech enhancement is not stable, which causes obvious speech distortions in the synthesized speech. Although System DCUnet, which adopts supervised speech enhancement, shows better ability of de-noising, it also suffers a lot from mispronunciation errors. Besides, there are also some noticable distortions in the generated speech. Note that for the proposed adversarial feature clustering method, we need extra clean speech data from another speaker as augmentation, but speech enhancement model is not needed.

\begin{table}
    \centering
        \caption{The performance for textual and acoustic noise.}
    \label{tab:founddata}
    \begin{tabular}{c|c |c || c| c| c}
        \hline
        Index  & CER (\%) & SNR &  MOS   & MCD & CGER \\ 
       \hline
        H          & \multirow{5}*{23.3} & \multirow{5}*{4 dB} & 0.78&  - & -\\
        Separabl~\cite{gurunath2019disentangling}   &  & & 2.35 & 8.70 & 3.67\%\\ 
        DCUnet~\cite{choi2019phase}     &  & & 3.23 & 5.32 & 2.11\% \\ 
		Adv-sen    &  & & 3.50 &6.51 & 0.88\%\\
        Adv-frame  &  & & 4.05 &5.03 & 0.23\%\\ 
        \hline
    \end{tabular}
    \vspace{-20pt}
\end{table}

With the help of another clean speech synthesis dataset, we conduct adversarial training to obtain noise-independent features in both sentence- and frame-level, named Adv-sen and Adv-frame respectively. For System Adv-sen, classification is conducted on the mean and variance of latent features to obtain sentence-level representation. We find although it can produce clean speech with good prosody, generated speech is not stable on pronunciations. With frame-level adversarial feature learning, System Adv-frame achieves the best performance among all systems. We assume the result benefits from two aspects: 1) the auxiliary dataset decreases the CER in the whole training texts and guides the model how to generate clean speech, 2) the adversarial feature learning can disentangle the noise information from speech, hence the control vector can directly control the generation process for producing clean speech.

\vspace{-1pt}
\section{Conclusions and Future Work}
\label{sec:conc}
This paper proposes an unsupervised clustering method to handle textual noise in found data, and an adversarial feature learning method to generate clean synthesized speech with noisy training speech. Experiment shows that the proposed methods are effective to build high-quality and stable seq2seq based speech synthesis model for noisy found data. Future work will try more robust methods to handle textual noise and for multi-speaker found data.

\bibliographystyle{IEEEtran}
\bibliography{main}

\begin{thebibliography}{10}
\providecommand{\url}[1]{#1}
\csname url@samestyle\endcsname
\providecommand{\newblock}{\relax}
\providecommand{\bibinfo}[2]{#2}
\providecommand{\BIBentrySTDinterwordspacing}{\spaceskip=0pt\relax}
\providecommand{\BIBentryALTinterwordstretchfactor}{4}
\providecommand{\BIBentryALTinterwordspacing}{\spaceskip=\fontdimen2\font plus
\BIBentryALTinterwordstretchfactor\fontdimen3\font minus
  \fontdimen4\font\relax}
\providecommand{\BIBforeignlanguage}[2]{{%
\expandafter\ifx\csname l@#1\endcsname\relax
\typeout{** WARNING: IEEEtran.bst: No hyphenation pattern has been}%
\typeout{** loaded for the language `#1'. Using the pattern for}%
\typeout{** the default language instead.}%
\else
\language=\csname l@#1\endcsname
\fi
#2}}
\providecommand{\BIBdecl}{\relax}
\BIBdecl

\bibitem{bahdanau2014neural}
D.~Bahdanau, K.~Cho, and Y.~Bengio, ``{Neural machine translation by jointly
  learning to align and translate},'' \emph{arXiv preprint arXiv:1409.0473},
  2014.

\bibitem{sutskever2014sequence}
I.~Sutskever, O.~Vinyals, and Q.~V. Le, ``Sequence to sequence learning with
  neural networks,'' in \emph{Proc. NIPS}, 2014, pp. 3104--3112.

\bibitem{wang2017tacotron}
Y.~Wang, R.~Skerry-Ryan, D.~Stanton, Y.~Wu, R.~J. Weiss, N.~Jaitly, Z.~Yang
  \emph{et~al.}, ``{Tacotron: Towards end-to-end speech synthesis},'' in
  \emph{Proc. INTERSPEECH}, 2017, pp. 4006--4010.

\bibitem{shen2017natural}
J.~Shen, R.~Pang, R.~J. Weiss, M.~Schuster, N.~Jaitly, Z.~Yang, Z.~Chen,
  Y.~Zhang, Y.~Wang, R.~Skerry-Ryan \emph{et~al.}, ``{Natural TTS synthesis by
  conditioning wavenet on mel spectrogram predictions},'' \emph{arXiv preprint
  arXiv:1712.05884}, 2017.

\bibitem{tachibana2017efficiently}
H.~Tachibana, K.~Uenoyama, and S.~Aihara, ``{Efficiently Trainable
  Text-to-Speech System Based on Deep Convolutional Networks with Guided
  Attention},'' in \emph{Proc. ICASSP}.\hskip 1em plus 0.5em minus 0.4em\relax
  IEEE, 2018, pp. 4784--4788.

\bibitem{li2019close}
N.~Li, S.~Liu, Y.~Liu, S.~Zhao, M.~Liu, and M.~Zhou, ``{Close to Human Quality
  TTS with Transformer},'' in \emph{Proc. AAAI}, 2019.

\bibitem{yang2020localness}
S.~Yang, H.~Lu, S.~Kang, L.~Xue, J.~Xiao, D.~Su, L.~Xie, and D.~Yu, ``On the
  localness modeling for the self-attention based end-to-end speech
  synthesis.'' \emph{Neural networks}, vol. 125, pp. 121--130, 2020.

\bibitem{valentini2016investigating}
C.~Valentini-Botinhao, X.~Wang, S.~Takaki, and J.~Yamagishi, ``Investigating
  rnn-based speech enhancement methods for noise-robust text-to-speech.'' in
  \emph{SSW}, 2016, pp. 146--152.

\bibitem{gurunath2019disentangling}
N.~Gurunath, S.~K. Rallabandi, and A.~Black, ``Disentangling speech and
  non-speech components for building robust acoustic models from found data,''
  \emph{arXiv preprint arXiv:1909.11727}, 2019.

\bibitem{hsu2019disentangling}
W.-N. Hsu, Y.~Zhang, R.~J. Weiss, Y.-A. Chung, Y.~Wang, Y.~Wu, and J.~Glass,
  ``Disentangling correlated speaker and noise for speech synthesis via data
  augmentation and adversarial factorization,'' in \emph{ICASSP 2019-2019 IEEE
  International Conference on Acoustics, Speech and Signal Processing
  (ICASSP)}.\hskip 1em plus 0.5em minus 0.4em\relax IEEE, 2019, pp. 5901--5905.

\bibitem{fong2019investigating}
J.~Fong, P.~O. Gallegos, Z.~Hodari, and S.~King, ``Investigating the robustness
  of sequence-to-sequence text-to-speech models to imperfectly-transcribed
  training data,'' in \emph{Proc. Interspeech}, 2019.

\bibitem{prenger2019waveglow}
R.~Prenger, R.~Valle, and B.~Catanzaro, ``Waveglow: A flow-based generative
  network for speech synthesis,'' in \emph{ICASSP 2019-2019 IEEE International
  Conference on Acoustics, Speech and Signal Processing (ICASSP)}.\hskip 1em
  plus 0.5em minus 0.4em\relax IEEE, 2019, pp. 3617--3621.

\bibitem{chorowski2015attention}
J.~K. Chorowski, D.~Bahdanau, D.~Serdyuk, K.~Cho, and Y.~Bengio,
  ``Attention-based models for speech recognition,'' in \emph{Proc. NPIS},
  2015, pp. 577--585.

\bibitem{battenberg2019location}
E.~Battenberg, R.~Skerry-Ryan, S.~Mariooryad, D.~Stanton, D.~Kao, M.~Shannon,
  and T.~Bagby, ``Location-relative attention mechanisms for robust long-form
  speech synthesis,'' \emph{arXiv preprint arXiv:1910.10288}, 2019.

\bibitem{van2017neural}
A.~van~den Oord, O.~Vinyals \emph{et~al.}, ``Neural discrete representation
  learning,'' in \emph{Advances in Neural Information Processing Systems},
  2017, pp. 6306--6315.

\bibitem{dunbar2019zero}
E.~Dunbar, R.~Algayres, J.~Karadayi, M.~Bernard, J.~Benjumea, X.-N. Cao,
  L.~Miskic, C.~Dugrain, L.~Ondel, A.~W. Black \emph{et~al.}, ``The zero
  resource speech challenge 2019: Tts without t,'' \emph{arXiv preprint
  arXiv:1904.11469}, 2019.

\bibitem{dilokthanakul2016deep}
N.~Dilokthanakul, P.~A. Mediano, M.~Garnelo, M.~C. Lee, H.~Salimbeni,
  K.~Arulkumaran, and M.~Shanahan, ``Deep unsupervised clustering with gaussian
  mixture variational autoencoders,'' \emph{arXiv preprint arXiv:1611.02648},
  2016.

\bibitem{sun2017unsupervised}
S.~Sun, B.~Zhang, L.~Xie, and Y.~Zhang, ``An unsupervised deep domain
  adaptation approach for robust speech recognition,'' \emph{Neurocomputing},
  vol. 257, pp. 79--87, 2017.

\bibitem{vincent2017analysis}
E.~Vincent, S.~Watanabe, A.~A. Nugraha, J.~Barker, and R.~Marxer, ``An analysis
  of environment, microphone and data simulation mismatches in robust speech
  recognition,'' \emph{Computer Speech \& Language}, vol.~46, pp. 535--557,
  2017.

\bibitem{choi2019phase}
H.-S. Choi, J.-H. Kim, J.~Huh, A.~Kim, J.-W. Ha, and K.~Lee, ``Phase-aware
  speech enhancement with deep complex u-net,'' \emph{arXiv preprint
  arXiv:1903.03107}, 2019.

\end{thebibliography}

\end{document}